\documentclass[aps,prl,showpacs,twocolumn,groupedaddress,amsmath,amssymb]{revtex4}
\usepackage{amssymb}
\usepackage{graphicx}
\usepackage{dcolumn}
\usepackage{bm}
\usepackage{color}
\usepackage{subfigure}
\usepackage{xspace}
\usepackage{textpos}
\usepackage[english]{babel}

\newcommand{\etapr}{\ensuremath{\eta^\prime}\xspace}

\newcommand{\un}[1]{\ensuremath{\,\mathrm{#1}}}
\def \mev        {\ensuremath{\un{MeV}\xspace}}

\def \gev        {\ensuremath{\un{GeV}\xspace}}
\def \gevc       {\ensuremath{\un{GeV}/c^2}\xspace}
\def \gevcc      {\ensuremath{\un{GeV}/c^2}\xspace}

\begin{document}

\begin{textblock*}{5cm}(130mm,-10mm)
  \flushright \textbf{BESIII-PUB-67}
\end{textblock*}

\title{\boldmath Observation of $\eta^{\prime}\to\pi^{+}\pi^{-}\pi^{+}\pi^{-}$
and $\eta^{\prime}\to\pi^{+}\pi^{-}\pi^{0}\pi^{0}$}

\author{
\small M.~Ablikim$^{1}$, M.~N.~Achasov$^{8,a}$, X.~C.~Ai$^{1}$,
O.~Albayrak$^{4}$, M.~Albrecht$^{3}$, D.~J.~Ambrose$^{41}$,
F.~F.~An$^{1}$, Q.~An$^{42}$, J.~Z.~Bai$^{1}$, R.~Baldini
Ferroli$^{19A}$
% ...
\\
\vspace{0.2cm}
(BESIII Collaboration)\\
\vspace{0.2cm} {\it
$^{1}$ Institute of High Energy Physics, Beijing 100049, People's Republic of China\\
% ...
$^{3}$ Bochum Ruhr-University, D-44780 Bochum, Germany\\
$^{4}$ Carnegie Mellon University, Pittsburgh, Pennsylvania 15213, USA\\
% ...
$^{8}$ G.I. Budker Institute of Nuclear Physics SB RAS (BINP), Novosibirsk 630090, Russia\\
% ...
$^{19}$ (A)INFN Laboratori Nazionali di Frascati, I-00044, Frascati, Italy; (B)INFN and University of Perugia, I-06100, Perugia, Italy\\
% ...
$^{41}$ University of Rochester, Rochester, New York 14627, USA\\
$^{42}$ University of Science and Technology of China, Hefei 230026, People's Republic of China\\
% ...
$^{a}$ Also at the Novosibirsk State University, Novosibirsk, 630090, Russia\\
% ...
} \vspace{0.4cm}
}

\vspace{0.4cm}

\begin{abstract}
  Using a sample of $1.3\times 10^9$ $J/\psi$ events collected with
  the BESIII detector, we report the first observation of
  $\eta^{\prime}\to\pi^{+}\pi^{-}\pi^{+}\pi^{-}$ and
  $\eta^{\prime}\to\pi^{+}\pi^{-}\pi^{0}\pi^{0}$. The measured
  branching fractions are
  $\mathcal{B}$($\eta^{\prime}\to\pi^{+}\pi^{-}\pi^{+}\pi^{-}$) =
  $(8.53\pm0.69({\rm stat.})\pm0.64({\rm syst.}))\times10^{-5}$ and
  $\mathcal{B}$($\eta^{\prime}\to\pi^{+}\pi^{-}\pi^{0}\pi^{0}$) =
  $(1.82\pm0.35({\rm stat.})\pm0.18({\rm syst.}))\times10^{-4}$, which
  are consistent with theoretical predictions based on a combination
  of chiral perturbation theory and vector-meson dominance.
\end{abstract}


\maketitle


The \etapr meson is much heavier than the Goldstone bosons of
broken chiral symmetry, and it has a special role in hadron physics
because of its interpretation as a singlet state arising due to the
axial $U(1)$ anomaly \cite{UA1,TH01}.  Discovered in
1964~\cite{intro2,intro3}, it remains a subject of extensive
theoretical studies aiming at extensions of chiral perturbation
theory \cite{CHPT1}.

New insight might be provided by the four-pion decays of \etapr.
The strong decays $\etapr \to \pi^+\pi^- \pi^{+(0)}\pi^{-(0)}$ are
not suppressed by approximate symmetries; they are expected to be
mediated by chiral anomalies, since an odd number (five) of
pseudoscalar particles are involved.  In particular, a contribution
from a new type of anomaly, the pentagon anomaly, might show up.
There should be also a significant contribution from the intermediate
state with two $\rho$ mesons. The four-pion
decays have not yet been observed, and the best upper limits until now come
from the CLEO collaboration:
$\mathcal{B}(\eta^{\prime}\to\pi^{+}\pi^{-}\pi^{+}\pi^{-}) <
2.4\times10^{-4}$ and $\mathcal{B} (\eta^{\prime}
\to\pi^{+}\pi^{-}\pi^{0}\pi^{0}) < 2.6 \times10^{-3}$ at the 90\%
confidence level (C.L.)~\cite{CLEO}.  Three decades ago,
a theoretical calculation
using the broken-SU$_6\times$O$_3$ quark
model~\cite{Parashar:1979js} yielded a branching ratio of $1.0\times
10^{-3}$ for $\mathcal{B}(\eta'\to \pi^+\pi^-
\pi^{+(0)}\pi^{-(0)})$.  For $\eta'\to \pi^+\pi^- \pi^+\pi^-$, this
value has already been excluded by the CLEO limit. Recently Guo,
Kubis and Wirzba~\cite{GuoFK}, using a combination of chiral
perturbation theory (ChPT) and a vector-meson dominance (VMD) model, obtained the
following prediction:
$\mathcal{B}(\eta^{\prime}\to\pi^{+}\pi^{-}\pi^{+}\pi^{-}) = (1.0
\pm 0.3)\times10^{-4}$ and
$\mathcal{B}(\eta^{\prime}\to\pi^{+}\pi^{-}\pi^{0}\pi^{0}) = (2.4
\pm 0.7)\times10^{-4}$. In this Letter, we report the first
observation of $\eta^{\prime}\to\pi^{+}\pi^{-}\pi^{+}\pi^{-}$ and
$\eta^{\prime}\to\pi^{+}\pi^{-}\pi^{0}\pi^{0}$ decays coming from
$J/\psi\rightarrow\gamma\eta^\prime$ radiative decay events using a
sample of $1.3\times 10^9$ $J/\psi$ events ($2.25\times 10^8$
events~\cite{data1} in 2009 and $1.09\times 10^9$ in
2012)~\cite{data2} taken at the center of mass energy of $3.097$ GeV
with the BESIII detector.

The BESIII detector is a magnetic spectrometer~\cite{bes3} located at
the Beijing Electron Position Collider (BEPCII), which is a
double-ring $e^+e^-$ collider with a design peak luminosity of
$10^{33} ~\rm{cm}^{-2}\rm{s}^{-1}$ at the center of mass energy of
$3.773\gev$. The cylindrical core of the BESIII detector consists of a
helium-based main drift chamber (MDC), a plastic scintillator
time-of-flight system (TOF), and a CsI(Tl) electromagnetic calorimeter
(EMC), which are all enclosed in a superconducting solenoidal magnet
providing a 1.0~T (0.9~T in 2012) magnetic field .  The solenoid is
supported by an octagonal flux-return yoke with resistive plate
counter muon identifier modules interleaved with steel. The acceptance
of charged particles and photons is 93\% over 4$\pi$ solid angle. The
charged-particle momentum resolution at $1\gevc$ is $0.5\%$, and the
$dE/dx$ resolution is 6\%.  The EMC measures photon energies with a
resolution of 2.5\% (5\%) at $1\gev$ in the barrel (endcaps).  The time
resolution of TOF is 80~ps in the barrel and 110~ps in the end caps.

Monte Carlo (MC) simulations are used to estimate backgrounds and
determine detection efficiencies.  Simulated events are processed
using \textsc{geant4}~\cite{geant4,geant4-bes}, where measured
detector resolutions are incorporated.

For $J/\psi\rightarrow\gamma\etapr$ with $\etapr \rightarrow
\pi^+\pi^-\pi^+\pi^-$, candidate events are required to have four
charged tracks and at least one photon. Each charged track,
reconstructed using hits in the MDC, is required to be in the polar
range $|\cos\theta| < 0.93$ and pass within 10 cm in the beam
direction and within 1 cm in the radial direction, with respect to the
interaction point.  For each charged track, the TOF and $dE/dx$
information are combined to form particle identification confidence
levels for the $\pi$, $K$, and $p$ hypotheses, and the particle type
with the highest C.L. is assigned to each track. At least two
oppositely charged tracks are required to be identified as
pions. Photon candidates, reconstructed by clustering EMC crystal
energies, must have at least $25\mev$ of energy for barrel showers
($|\cos\theta|<0.8$), or $50\mev$ for endcap showers ($0.86 <
|\cos\theta| < 0.92$). To exclude showers from charged particles, the
angle between the nearest charged track and the shower must be greater
than 10$^{\circ}$.  Further, EMC cluster timing requirements are used
to suppress electronic noise and energy deposits unrelated to the
event.

\begin{figure}
      \includegraphics[width=6.5cm,height=5.6cm]{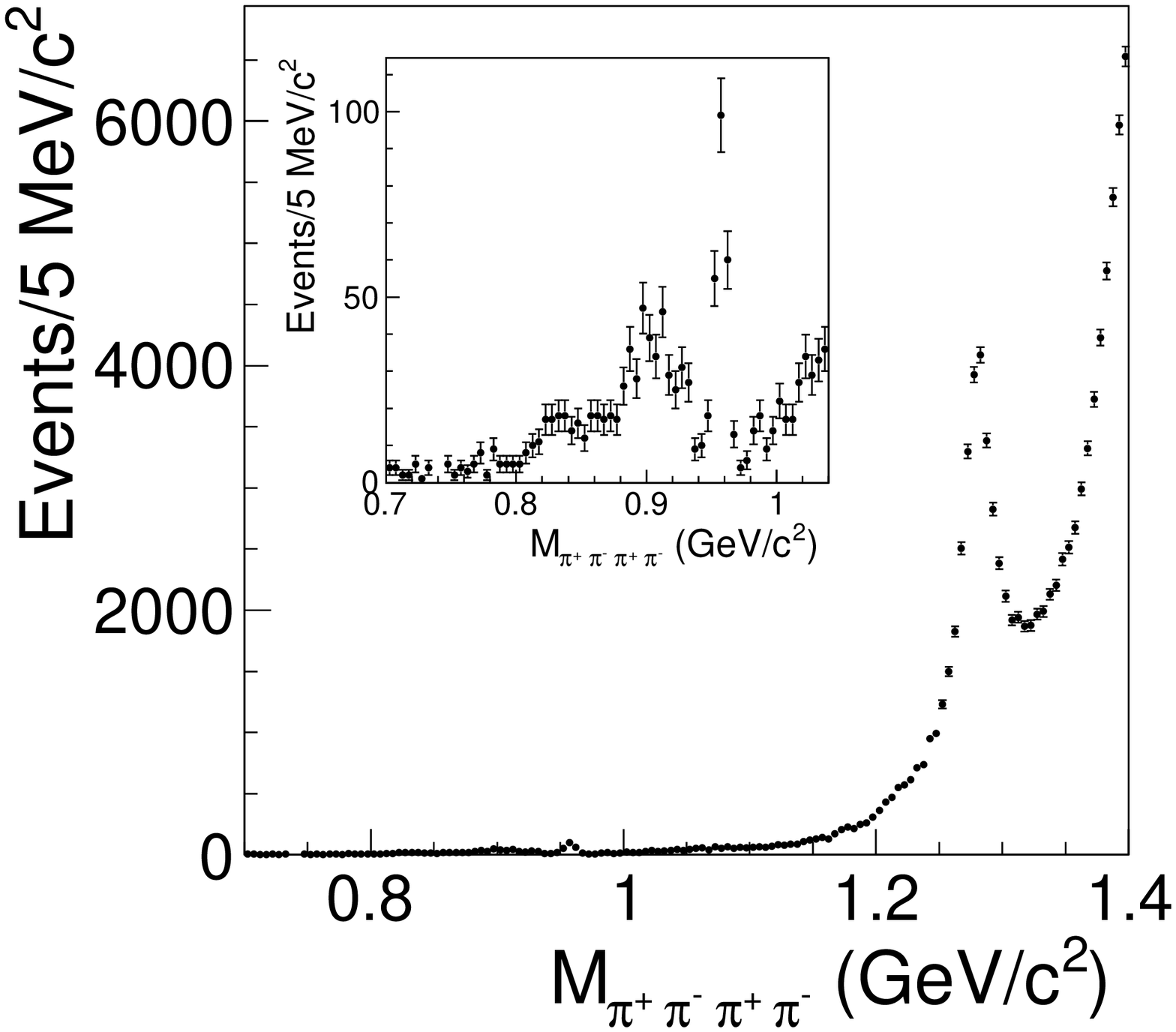}
\put(-35,125){(a)}

      \includegraphics[width=6.5cm,height=5.6cm]{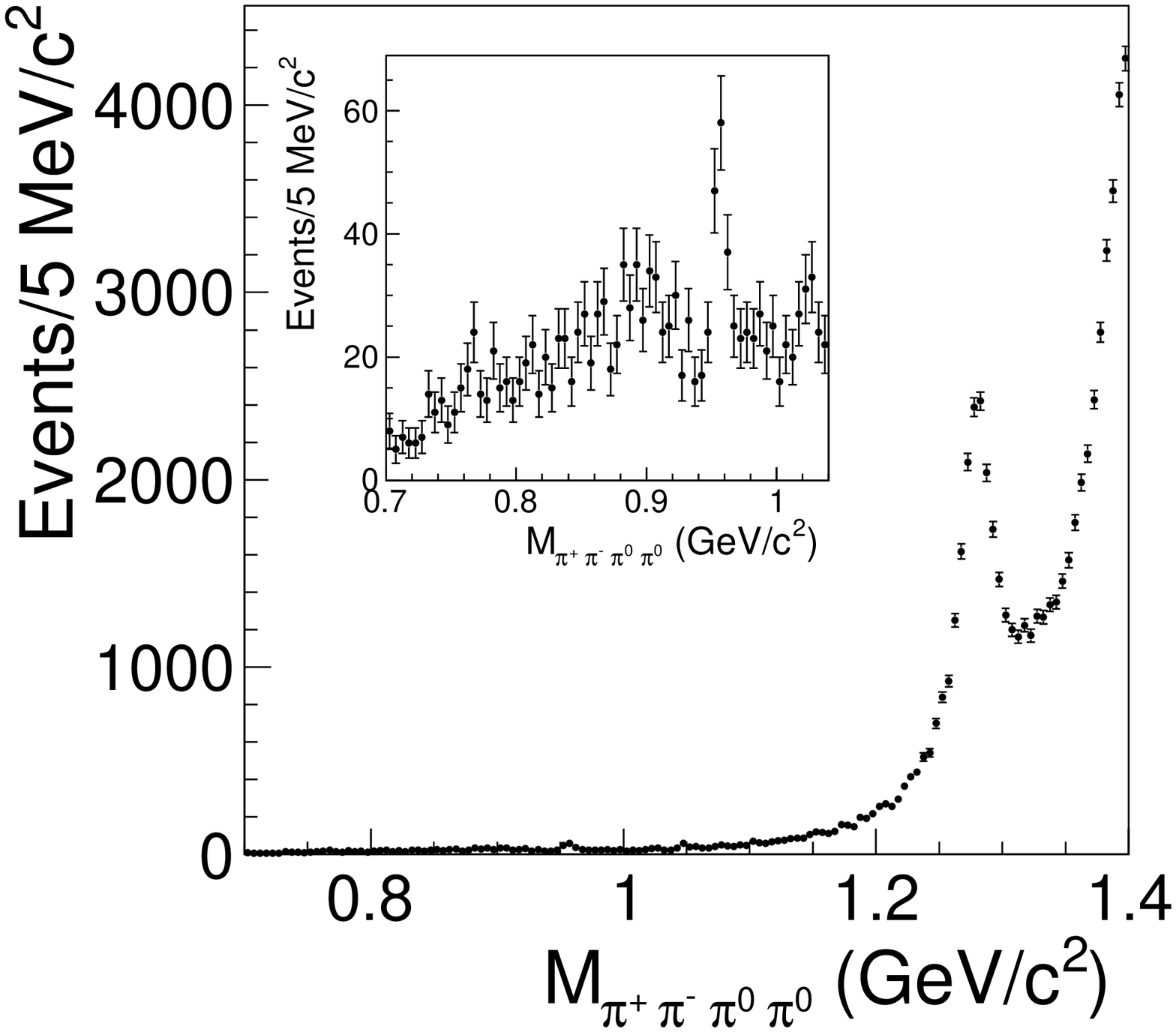}
\put(-35,125){(b)}
\caption{ The invariant mass distributions of (a)
  $\pi^+\pi^-\pi^+\pi^-$ and (b) $\pi^+\pi^-\pi^0\pi^0$ after the
  final selection. The inserts are for the mass spectra around
 the  $\eta^\prime$ mass region.
\label{m4pi}}
\end{figure}

Next a four-constraint (4C) kinematic fit imposing energy-momentum
conservation is performed under the $\gamma\pi^+\pi^-\pi^+\pi^-$
hypothesis, and a loose requirement of $\chi^2_{\rm 4C} < 35$ is
imposed. If there is more than one photon candidate in an event, the
combination with the smallest $\chi^2_{\rm 4C}$ is retained, and its
$\chi^2_{\rm 4C}$ is required to be less than that for the
$\gamma\gamma\pi^+\pi^-\pi^+\pi^-$ hypothesis.  The
$\pi^+\pi^-\pi^+\pi^-$ invariant mass spectrum for the selected
events is shown in Fig.~\ref{m4pi}(a), where an $\eta^\prime$ peak
is clearly observed in the inset plot.

To ensure that the $\etapr$ peak is not from background, a study was
performed with a MC sample of 1 billion $J/\psi$ events generated
with the $\sc Lund$ model~\cite{lund}. The results indicate that the
enhancement below the $\etapr$ peak in Fig.~\ref{m4pi}(a) is from
the background channel $\etapr \rightarrow \pi^+\pi^-\eta$ with
$\eta\rightarrow\gamma\pi^+\pi^-$, while the background in the mass
region above $1\gevcc$ is mainly from $\etapr \rightarrow
\pi^+\pi^-e^+e^-$. Other background channels are $J/\psi \to \gamma
f_2(1270), f_2(1270)\rightarrow\pi^+\pi^-\pi^+\pi^-$ and
non-resonant $J/\psi\rightarrow\gamma\pi^+\pi^-\pi^+\pi^-$. However,
none of these background sources produces a peak in the
$\pi^+\pi^-\pi^+\pi^-$ invariant mass spectrum near
  the \etapr mass.

\begin{figure}
      \includegraphics[width=6.5cm,height=6.5cm]{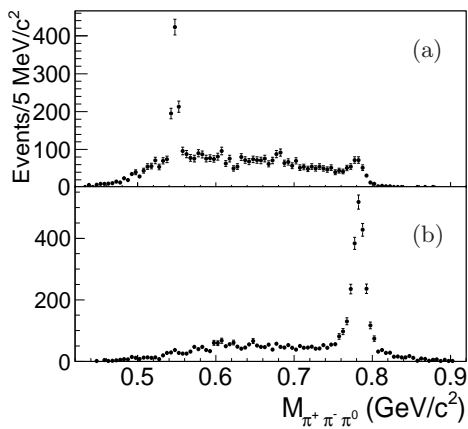}
\put(-30,160){(a)}
\put(-30,90){(b)}

    \caption{ The $\pi^+\pi^-\pi^0$ invariant mass distributions for
      the combinations closest to (a) $m_\eta$ and (b) $m_\omega$.
      \label{m3pi}}
\end{figure}

For $J/\psi\rightarrow\gamma\eta^\prime$ with
$\eta^\prime\rightarrow \pi^+\pi^-\pi^0\pi^0$, candidate events must
have two charged tracks with zero net charge, that are identified as
pions, and at least five photons.  One-constraint (1C) kinematic
fits are performed on the $\pi^{0}$ candidates reconstructed from
photon pairs with the invariant mass of the two photons being
constrained to the $\pi^0$ mass, and $\chi^{2}_{\rm
1C}(\gamma\gamma)<50$ is required.  Then a six-constraint (6C)
kinematic fit (two $\pi^{0}$ masses are also constrained) is
performed under the hypothesis of
$J/\psi\rightarrow\gamma\pi^{+}\pi^{-}\pi^{0}\pi^{0}$. For events
with more than two $\pi^0$ candidates, the combination with the
smallest $\chi^{2}_{\rm 6C}$ is retained. A rather loose criterion
of  $\chi^2 < 35$ is required to exclude events with a kinematics
incompatible with the signal hypothesis. To reject background from
events with six photons in the final state, $\chi^2_{\rm 6C}$ is
required to be less than that for the
$\gamma\gamma\pi^+\pi^-\pi^0\pi^0$ hypothesis. After this selection,
Figs.~\ref{m3pi}(a) and (b) show the invariant mass of the
$\pi^+\pi^-\pi^0$ combination closest to the nominal $\eta$ or
$\omega$ mass (denoted as $m_{\eta}$ and $m_{\omega}$),
respectively. Significant $\eta$ and $\omega$ peaks are seen. These
backgrounds are suppressed by rejecting events with
$|M_{\pi^{+}\pi^{-}\pi^{0}}-m_{\eta}| <0.02 \gevcc$ or
$|M_{\pi^{+}\pi^{-}\pi^{0}}-m_{\omega}| < 0.02\gevcc$.

After the above selection, Fig.~\ref{m4pi}(b) shows the
$\pi^+\pi^-\pi^0\pi^0$ invariant mass distribution, where an
$\eta^\prime$ peak is very clear.  With the MC sample of 1 billion
$J/\psi$ events, the same study for
$\eta^\prime\rightarrow\pi^+\pi^-\pi^0\pi^0$ was also performed to
investigate possible background events, and the main backgrounds
were found to come from: (1) $\eta^{\prime}\to\pi^{+}\pi^{-}\eta$,
$\eta\to\pi^{0}\pi^{0}\pi^{0}$, (2)
$\eta^{\prime}\to\pi^{0}\pi^{0}\eta$, $\eta\to\gamma\pi^{+}\pi^{-}$,
(3) $\eta^{\prime}\to\gamma\omega$,
$\omega\to\pi^{+}\pi^{-}\pi^{0}$, and (4) $J/\psi \to \gamma
f_2(1270), f_2(1270)\rightarrow\pi^+\pi^-\pi^0\pi^0$ and (5)
non-resonant $J/\psi\rightarrow\gamma\pi^+\pi^-\pi^0\pi^0$. None of
these possible background channels contribute to the $\eta^\prime$
peak.

The signal yields are obtained from extended unbinned maximum
likelihood fits to the $ \pi^+\pi^-\pi^{+(0)}\pi^{-(0)}$ invariant
mass distributions. The total probability density function (PDF)
consists of a signal and various background contributions.  The
signal component is modeled as the MC simulated signal shape
convoluted with a Gaussian function to account for the difference in
the mass resolution observed between data and MC simulation. For
this analysis, MC simulation indicates that the mass resolution has
almost no change for the two data sets taken in 2009 and 2012,
respectively. The background components considered are subdivided
into three classes: (i) the shapes of those background events that
contribute to a structure in $M_{\pi^+\pi^-\pi^+\pi^-}$ [e.g.,
$\eta^\prime\rightarrow\pi^+\pi^-\eta$ with
$\eta\rightarrow\gamma\pi^+\pi^-$ and
$\eta^\prime\rightarrow\pi^+\pi^- e^+e^-$] or
$M_{\pi^+\pi^-\pi^0\pi^0}$ [e.g.,
$\eta^{\prime}\to\pi^{+}\pi^{-}\eta$ with
$\eta\to\pi^{0}\pi^{0}\pi^{0}$ and
$\eta^{\prime}\to\pi^{0}\pi^{0}\eta$ with
$\eta\to\gamma\pi^{+}\pi^{-}$, as well as
$\eta^{\prime}\to\gamma\omega$ with
$\omega\to\pi^{+}\pi^{-}\pi^{0}$] are taken from the dedicated MC
simulations; (ii) the tail of the resonance $f_2(1270)$ from $J/\psi
\to \gamma f_2(1270)$ is parameterized with a Breit-Wigner function
convoluted with a Gaussian for the mass resolution from the
simulation; (iii) $J/\psi\rightarrow\gamma\pi^+\pi^-\pi^+\pi^-$
($J/\psi\rightarrow\gamma\pi^+\pi^-\pi^0\pi^0$) phase space is also
described with the MC simulation shape.  In the fit to data, the
mass and width of $f_2(1270)$ are fixed to the values in the
PDG~\cite{pdg}, while the magnitudes of different components are
left free
in the fit to account for the
uncertainties of the branching fractions of
$J/\psi\rightarrow\gamma\eta^\prime$ and other intermediate decays
(e.g., $\eta^\prime\rightarrow\pi^+\pi^-\eta$,
$\eta^\prime\rightarrow\pi^0\pi^0\eta$, and
$\eta\rightarrow\gamma\pi^+\pi^-$).

\begin{figure}

    \includegraphics[width=6.5cm,height=5.6cm]{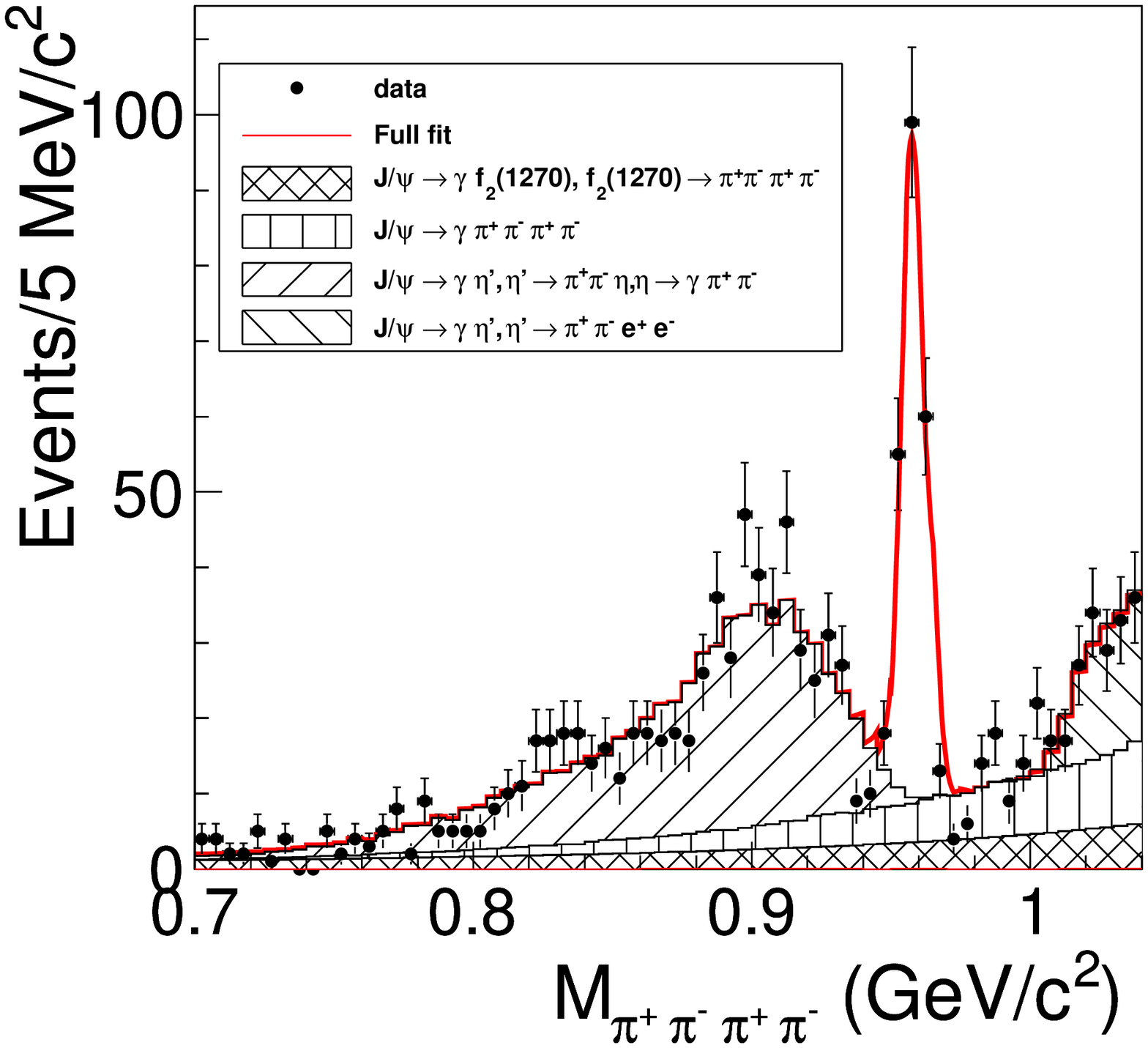}
\put(-30,125){(a)}

    \includegraphics[width=6.5cm,height=5.6cm]{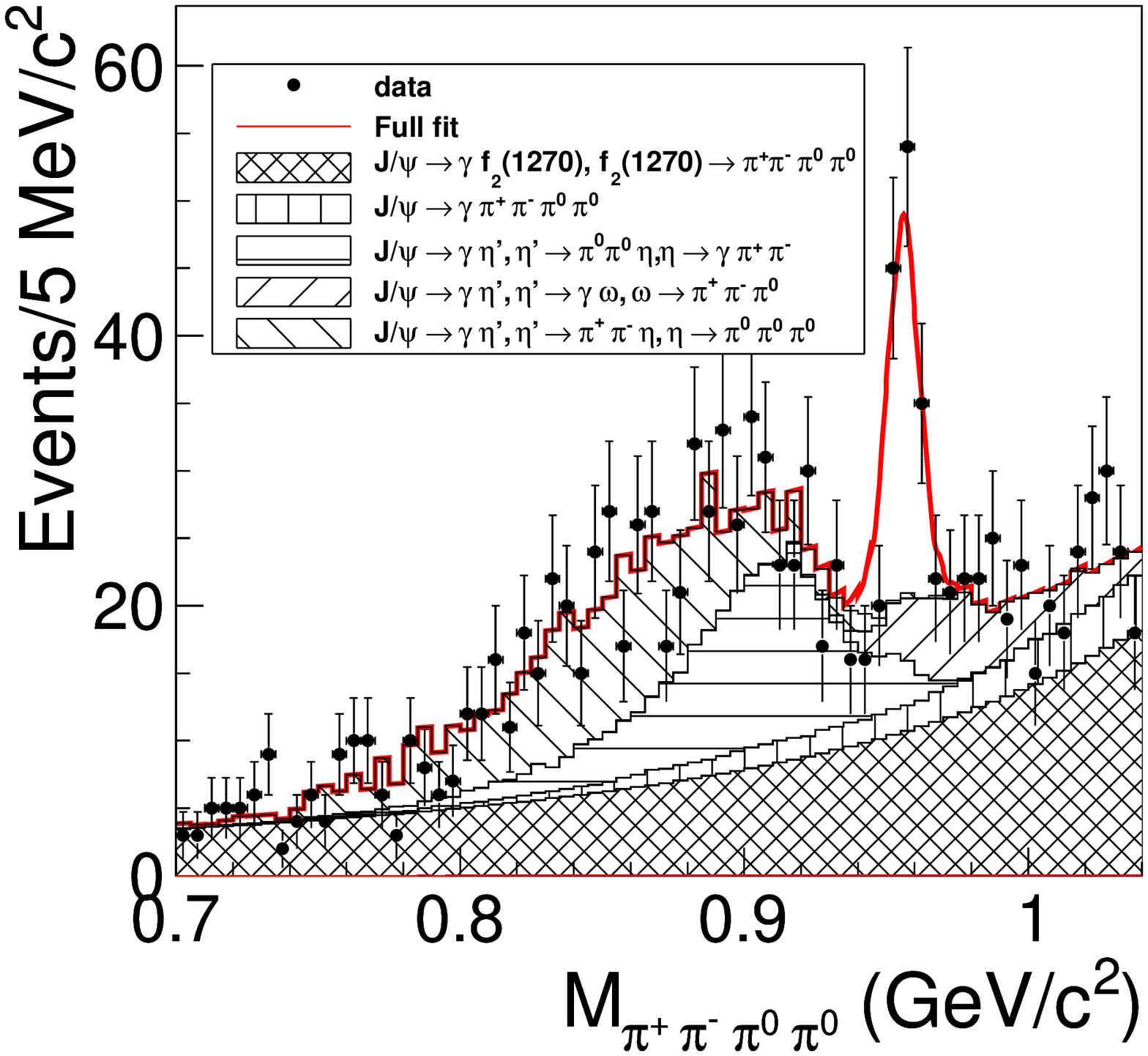}
\put(-30,125){(b)}
    \caption{ Results of the fits to (a) $M_{\pi^+\pi^-\pi^+\pi^-}$ and
(b) $M_{\pi^+\pi^-\pi^0\pi^0}$, where the background contributions are displayed
as the hatched histograms.
\label{m4pi_fit}}
\end{figure}

The projections of the fit to $M_{\pi^+\pi^-\pi^{+(0)}\pi^{-(0)}}$
in the $\eta^\prime$ mass region are shown in
Figs.~\ref{m4pi_fit}(a) and (b), where the shape of the sum of
signal and background shapes is in good agreement with data. We
obtain $199\pm 16$ $\eta^\prime\rightarrow \pi^+\pi^-\pi^+\pi^-$
events with a statistical significance of 18$\sigma$ and $84 \pm 16$
$\eta^{\prime}\rightarrow \pi^+\pi^-\pi^0\pi^0$ events with a
statistical significance of 5$\sigma$.  The statistical significance
is determined by the change of the log-likelihood value and the
number of degree of freedom in the fit with and without the
$\eta^\prime$ signal.

\begin{figure}
      \includegraphics[width=6.0cm,height=5.6cm]{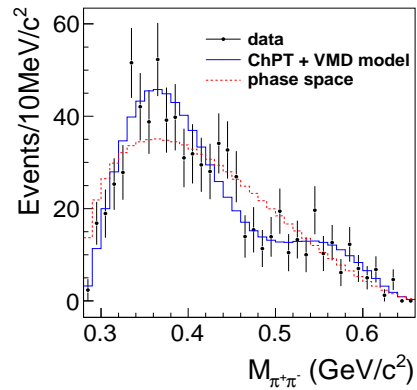}
   \caption{ The comparison of $M_{\pi^{+}\pi^{-}}$ (four entries per event)
between data and two different models, where the dots with error
bars are for the background-subtracted data, the solid line is for
the ChPT and VMD model, and the dashed line is for the phase space.
   \label{m2pi}}
   \end{figure}

In order to compute the branching fractions, the signal efficiencies
for the selection criteria described above are estimated with the MC
simulation. To ensure a good description of data, in addition to the
phase space events, we also produced a signal MC sample in which the
signal simulation is modeled as the decay amplitudes in
Ref.~\cite{GuoFK} based on the ChPT and VMD model. For
$\eta^\prime\rightarrow\pi^+\pi^-\pi^+\pi^-$, we divide each of
$M_{\pi^+\pi^-}$ combination into 38 bins in the region of
[0.28, 0.66] GeV/$c^{2}$. With the same procedure as described above, the number
of the $\eta^\prime$ events in each bin can be obtained by fitting
the $\pi^+\pi^-\pi^+\pi^-$ mass spectrum in this bin, and then the
background-subtracted $M_{\pi^+\pi^-}$ is obtained as shown in Fig.~\ref{m2pi}
(four entries per event), where the errors are statistical only.
The comparison of $M_{\pi^+\pi^-}$ between data and two different
models displayed in Fig.~\ref{m2pi} indicates that the ChPT and VMD model
could provide a more reasonable description of data than the phase
space events. Therefore the simulation events generated with the
ChPT and VMD model are applied to determine the detection efficiency
for $\eta^\prime\rightarrow\pi^+\pi^-\pi^{+(0)}\pi^{-(0)}$ decays.
Table~\ref{sum_br} lists all the information
used for the branching fraction measurements.


\begin{table}[htpb]
\begin{center}
\caption{Signal yields, detection efficiencies and the product branching
fractions of $J/\psi \rightarrow \gamma \eta^{\prime}$, $\eta^\prime\rightarrow \pi^+\pi^-\pi^{+(0)}\pi^{-(0)}$.
The first errors are statistical and the second systematic.}\label{sum_br}
\begin{tabular}{lccc}
\hline\hline
Mode &  Yield &  $\varepsilon$ (\%)& Branching fraction\\ \hline
$\eta^\prime\rightarrow\pi^{+}\pi^{-}\pi^{+}\pi^{-}$
&  $199\pm 16$  & 34.5 & $(4.40\pm0.35\pm0.30)\times10^{-7}$\\
$\eta^\prime\rightarrow\pi^{+}\pi^{-}\pi^{0}\pi^{0}$& $84\pm 16$ &
7.0
 & $(9.38\pm1.79\pm0.89)\times10^{-7}$\\
\hline
\hline
\end{tabular}
\end{center}
\end{table}

Sources of systematic errors and their corresponding contributions to
the measurement of the branching fractions are summarized in
Table~\ref{sumerr}.
The uncertainties in MDC tracking and photon
detection have been studied with the high purity control sample of
$J/\psi\rightarrow \rho\pi$ for two data sets.
The differences in the detection
efficiencies between data and MC simulation are less than 1\% per
charged track and 1\% per photon, which are taken as the systematic errors.
Similarly, to estimate the error related to the pion identification,
the pion identification efficiency has been studied using a clean sample of
$J/\psi\rightarrow \rho\pi$, and
the data is found to be in
agreement with MC simulation within 1\%.
For
$\eta^\prime\rightarrow\pi^+\pi^-\pi^+\pi^-$, at least one $\pi^+$ and
one $\pi^-$ are required to be identified, and the error from this
source is calculated to be 0.6\%, while 2\% is assigned for
$\eta^\prime\rightarrow\pi^+\pi^-\pi^0\pi^0$ because both charged
tracks are required to be identified as pions.  The uncertainty
arising from the $\omega$ ($\eta$) veto is estimated by varying the
requirements from $|M_{\pi^{+}\pi^{-}\pi^{0}}-m_{\eta} /m_{\omega}|
>0.02\gevcc$ to $|M_{\pi^{+}\pi^{-}\pi^{0}}-m_{\eta}/m_{\omega}|
>0.018\gevcc$ in the event selection.

The uncertainty associated with the kinematic fit comes from the
inconsistency between data and MC simulation of the track parameters
and the error matrices.  In this analysis the uncertainties arising
from the kinematic fit are estimated by using $J/\psi\to\phi\eta$
events with $\phi\to K^+ K^-$ and
$\eta\to\pi^{+(0)}\pi^{-(0)}\pi^0$, which have a topology similar to
the decay channels of interest. A sample is selected without a
kinematic fit. The event selection for charged tracks and photons
are the same as the two decays studied in this analysis. Each
charged track is identified as a kaon or a pion. Then a 4C kinematic
fit is performed for the candidates of $J/\psi\rightarrow\phi\eta$,
$\eta\rightarrow\pi^+\pi^-\pi^0$; and a 7C kinematic fit for
$J/\psi\rightarrow\phi\eta$, $\eta\rightarrow\pi^0\pi^0\pi^0$ by
constraining the $\gamma\gamma$ invariant mass to be the $\pi^0$
mass.  The efficiencies for $\chi^2<35$ are obtained by comparing
the number of signal events with and without the 4C (7C) kinematic
fit performed for data and MC simulation separately. The data-MC
differences shown in Table~\ref{sumerr} are taken as the systematic
errors from this source.

Background events whose distributions peak either below (e.g.,
$\eta^\prime\rightarrow \pi^+\pi^-\eta $) or just above the
$\eta^\prime$ peak (e.g., $\eta^\prime\rightarrow\gamma\omega$) may
alter the signal yield. We performed an alternative fit by fixing
these contributions according to the branching fractions of
$J/\psi\rightarrow \gamma \eta^\prime$ and the cascade decays and
found the impact on the signal yield is small. The uncertainty
associated with the smooth background functions, including the phase
space shape and the tail of $f_2(1270)$, is evaluated by replacing
them with a second order polynomial, and the uncertainties of 2.1\%
and 3.5\% are due to the yield difference with respect to the nominal
fit. The uncertainties due to the fit range are considered by varying
the fit ranges, and the difference of the results are 2.1\% and 3.8\%.
The uncertainties due to the MC model are estimated with MC samples in
which the signal simulation is modeled according to the decay
amplitudes in Ref.~\cite{GuoFK} and a phase space distribution, and
the differences are $1.4\%$ and $4.5\%$, respectively.

The branching fractions for $J/\psi\rightarrow\gamma\eta^\prime$ and
$\pi^0\rightarrow\gamma\gamma$ decays are taken from the world average
values~\cite{pdg}, and the uncertainties on these branching fractions are
taken as the associated systematic uncertainty in our measurements.

All the above contributions and the uncertainty from the number of
$J/\psi$ events~\cite{data2} are summarized in Table~\ref{sumerr},
where the total systematic uncertainty is given by the quadratic sum
of the individual errors, assuming all sources to be independent.

\begin{table}[htpb]
\begin{center}
  \caption{Summary of the systematic uncertainties in the branching
    fractions (in \%). In the calculation of the product branching
    fractions of $J/\psi \rightarrow \gamma
    \eta^{\prime},\eta^{\prime} \rightarrow
    \pi^{+}\pi^{-}\pi^{+(0)}\pi^{-(0)}$, the uncertainty of
    $\mathcal{B}(J/\psi\rightarrow\gamma\eta^\prime)$ is not
    included.}\label{system}
\label{sumerr}
\begin{tabular}{l c c}
\hline\hline
Sources & $\eta^{\prime}\to\pi^{+}\pi^{-}\pi^{+}\pi^{-}$ &
$\eta^{\prime}\to\pi^{+}\pi^{-}\pi^{0}\pi^{0}$ \\
\hline
MDC tracking & 4.0 & 2.0 \\
Photon detection & 1.0 & 5.0 \\
Particle identification & 0.6 & 2.0 \\
$\eta$ ($\omega$) veto & - & 2.1 \\
4C/6C kinematic fit & 4.4 & 2.1 \\
Continuous BG shape & 2.1 & 3.5 \\
Fit range & 2.1 & 3.8 \\
MC model & 1.4 & 4.5 \\

 $\mathcal{B}(J/\psi\rightarrow\gamma\eta^\prime)$ & 2.9 & 2.9 \\
 $\mathcal{B}(\pi^0\rightarrow\gamma\gamma)$ & - & 0.1 \\
Number of $J/\psi$ events & 0.8 & 0.8 \\\hline
Total & 7.5 & 9.9 \\
\hline\hline
\end{tabular}
\end{center}
\end{table}

In summary, based on a sample of 1.3 billion $J/\psi$ events taken
with the BESIII detector, we observe the decay modes
$\eta^{\prime}\to\pi^{+}\pi^{-}\pi^{+}\pi^{-}$ and
$\eta^{\prime}\to\pi^{+}\pi^{-}\pi^{0}\pi^{0}$ with a statistical
significance of 18$\sigma$ and 5$\sigma$, respectively, and measure
their product branching fractions: $\mathcal{B}(J/\psi \rightarrow
\gamma \eta^{\prime}) \cdot
\mathcal{B}(\eta^{\prime}\to\pi^{+}\pi^{-}\pi^{+}\pi^{-}) = (4.40\pm
0.35({\rm stat.})\pm 0.30({\rm syst.}))\times10^{-7}$ and
$\mathcal{B}(J/\psi \rightarrow \gamma \eta^{\prime}) \cdot
\mathcal{B}(\eta^{\prime}\to\pi^{+}\pi^{-}\pi^{0}\pi^{0}) = (9.38\pm
1.79({\rm stat.})\pm 0.89({\rm syst.}))\times10^{-7}$. Using the PDG
world average value of
$\mathcal{B}(J/\psi\rightarrow\gamma\eta^\prime)$~\cite{pdg}, the
branching fractions of
$\eta^{\prime}\to\pi^{+}\pi^{-}\pi^{+(0)}\pi^{-(0)}$ are determined
to be $\mathcal{B}(\eta^{\prime}\to\pi^{+}\pi^{-}\pi^{+}\pi^{-}) =
(8.53\pm0.69({\rm stat.})\pm0.64({\rm syst.}))\times10^{-5}$ and
$\mathcal{B}(\eta^{\prime}\to\pi^{+}\pi^{-}\pi^{0}\pi^{0}) =
(1.82\pm0.35({\rm stat.})\pm0.18({\rm syst.}))\times10^{-4}$, which
are consistent with the theoretical predictions based on a
combination of chiral perturbation theory and vector-meson
dominance, but not with the broken-SU$_6\times$O$_3$ quark
model~\cite{Parashar:1979js}.

The BESIII collaboration thanks the staff of BEPCII and the
computing center for their strong support. This work is supported in
part by the Ministry of Science and Technology of China under
Contract No. 2009CB825200; Joint Funds of the National Natural
Science Foundation of China under Contracts Nos. 11079008, 11179007,
U1232101, U1232107, U1332201; National Natural Science Foundation of
China (NSFC) under Contracts Nos. 10625524, 10821063, 10825524,
10835001, 10935007, 11125525, 11235011,11175189; the Chinese Academy
of Sciences (CAS) Large-Scale Scientific Facility Program; CAS under
Contracts Nos. KJCX2-YW-N29, KJCX2-YW-N45; 100 Talents Program of
CAS; German Research Foundation DFG under Contract No. Collaborative
Research Center CRC-1044; Istituto Nazionale di Fisica Nucleare,
Italy; Ministry of Development of Turkey under Contract No.
DPT2006K-120470; U. S. Department of Energy under Contracts Nos.
DE-FG02-04ER41291, DE-FG02-05ER41374, DE-FG02-94ER40823,
DESC0010118; U.S. National Science Foundation; University of
Groningen (RuG) and the Helmholtzzentrum fuer Schwerionenforschung
GmbH (GSI), Darmstadt; WCU Program of National Research Foundation
of Korea under Contract No. R32-2008-000-10155-0.

\end{document}